\documentclass[twocolumn]{aastex63}
\usepackage{apjfonts}

\received{2019 March 22}
\revised{2020 February 16}
\accepted{2020 March 6}

\shorttitle{Time Variability of Nonthermal X-ray Stripes in Tycho's SNR}
\shortauthors{Okuno et al.}

\begin{document}

\title{Time Variability of Nonthermal X-ray Stripes in Tycho's Supernova Remnant with Chandra}

\correspondingauthor{Tomoyuki Okuno}
\email{okuno.tomoyuki.32r@kyoto-u.jp}

\correspondingauthor{Takaaki Tanaka}
\email{ttanaka@cr.scphys.kyoto-u.ac.jp}

\author{Tomoyuki Okuno}
\affil{Department of Physics, Kyoto University, Kitashirakawa Oiwake-cho, Sakyo, Kyoto, Kyoto 606-8502, Japan}

\author{Takaaki Tanaka}
\affiliation{Department of Physics, Kyoto University, Kitashirakawa Oiwake-cho, Sakyo, Kyoto, Kyoto 606-8502, Japan}

\author{Hiroyuki Uchida}
\affiliation{Department of Physics, Kyoto University, Kitashirakawa Oiwake-cho, Sakyo, Kyoto, Kyoto 606-8502, Japan}

\author{Felix A. Aharonian}
\affil{Dublin Institute for Advanced Studies, 31 Fitzwilliam Place, Dublin 2, Ireland}
\affil{Max-Planck-Institut f\"ur Kernphysik, PO Box 103980, 69029 Heidelberg, Germany}

\author{Yasunobu Uchiyama}
\affiliation{Department of Physics, Rikkyo University, 3-34-1 Nishi Ikebukuro, Toshima-ku, Tokyo 171-8501, Japan}

\author{Takeshi~Go Tsuru}
\affiliation{Department of Physics, Kyoto University, Kitashirakawa Oiwake-cho, Sakyo, Kyoto, Kyoto 606-8502, Japan}

\author{Masamune Matsuda}
\affiliation{Department of Physics, Kyoto University, Kitashirakawa Oiwake-cho, Sakyo, Kyoto, Kyoto 606-8502, Japan}


\begin{abstract}
Analyzing Chandra data of Tycho's supernova remnant (SNR) taken in 2000, 2003, 2007, 2009, and 2015, we search for time variable features of 
synchrotron X-rays in the southwestern part of the SNR, where stripe structures of hard X-ray emission were previous found. 
By comparing X-ray images obtained at each epoch, we discover a knot-like structure in the northernmost part of the stripe region became brighter particularly in 2015. 
We also find a bright filamentary structure gradually became fainter and narrower as it moved outward. 
Our spectral analysis reveal that not only the nonthermal X-ray flux but also the photon indices of the knot-like structure change from year to year. 
During the period from 2000 to 2015, the small knot shows brightening of $\sim 70\%$ and hardening of $\Delta \Gamma \sim 0.45$. 
The time variability can be explained if the magnetic field is amplified to $\sim 100~\mathrm{\mu G}$ and/or if magnetic turbulence significantly changes with time. 
\end{abstract}


\section{Introduction}\label{sec:introduction}
Tycho's supernova remnant (Tycho's SNR, a.k.a. G120.1+1.4) is a Galactic supernova remnant whose supernova was recorded by Tycho Brahe in 1572.
Tycho's SNR has been thought to be of Type Ia origin from his records \citep[e.g.,][]{Baade1945,Ruiz-Lapuente2004}.
\cite{Krause2008} supported that Tycho's SNR resulted from a typical Type Ia supernova by performing the spectroscopy of the scattered-light echo.
The distance to Tycho's SNR is still uncertain, but most estimations agree on 2--4~kpc as reviewed by \cite{Hayato2010}.
For example, CO observations suggest that molecular clouds at $\sim 2.5~\mathrm{kpc}$ associate with Tycho's SNR \citep[e.g.,][]{Lee2004,Zhou2016,Chen2017}, although \cite{Tian2011} claimed Tycho's SNR has no associated molecular cloud.

Chandra, with its superb angular resolution, resolved thermal and nonthermal emission of Tycho's SNR into inner ejecta clumps and thin outer rims, respectively \citep[e.g.,][]{Hwang2002}.
Thin filamentary nonthermal emission is a common feature among young SNRs, suggesting strong magnetic field in the downstream region \citep[e.g.,][]{Bamba2005}.
In the Tycho's SNR case, the pre-shock magnetic field is estimated to be $\sim 30~\mathrm{\mu G}$ based on the width of the rim \citep[e.g.,][]{Cassam-Chenai2007}.
On the other hand, inner nonthermal X-ray stripes are unique features discovered so far only in Tycho's SNR. 
\cite{Eriksen2011} found that the bright western stripes and the faint southern stripes have relatively hard emission, and suggested magnetic field amplification in these stripes.
\cite{Bykov2011} reproduced the X-ray stripes with a nonlinear diffusive shock acceleration theory which includes magnetic field amplification from cosmic ray particle current-driven instability although 
the origin of the stripe structures is still under debate. 

If magnetic field amplification plays a role in generation of the stripes, one can expect time-variable synchrotron X-rays in the stripe regions due to fast acceleration and synchrotron cooling by the amplified magnetic field. 
Time-variable synchrotron X-rays were indeed discovered in the SNR RX~J1713.7$-$3946 by \cite{Uchiyama2007}. 
Similar time-variable features were found also in Cassiopeia A \citep{Uchiyama2008} and recently in G330.2+1.0 \citep{Borkowski2018}. 
 
In order to investigate the origin of the nonthermal stripes, we search for time variability of synchrotron X-rays in Tycho's SNR.
Tycho's SNR has been observed sufficiently by Chandra ACIS in 2000, 2003, 2007, 2009, and 2015.
We describe details of the data sets we used and data reduction procedures in section~\ref{sec:observations}.
We then perform imaging and spectral analyses in section~\ref{sec:analysis}, and then discuss our results in section~\ref{sec:discussion}.
Throughout this paper, quoted errors are all $1\sigma$ confidence intervals.

\section{Observations and data reduction}\label{sec:observations}
Tycho's SNR has been observed with the Advanced CCD Imaging Spectrometer \citep[ACIS; e.g.,][]{Garmire2003} aboard Chandra  in 2000, 2003, 2007, 2009, and 2015.
All the observations except for that in 2000 used the ACIS-I array, which consists of four front-illuminated (FI) CCD chips, and covered the entire part of Tycho's SNR with almost the same aim points and roll angles.
During the observation in 2000, Tycho's SNR was focused on the back-illuminated CCD chip, ACIS-S3. 

The data we analyzed are summarized in table \ref{tab:tycho_observations}.
We first reprocessed and screened the data with the standard criteria using the {\tt chandra\_repro} task in the analysis software package, CIAO version 4.9\footnote{\url{http://cxc.cfa.harvard.edu/ciao/}}, with the calibration data from CALDB version 4.7.8\footnote{\url{http://cxc.harvard.edu/caldb/downloads/Release_notes/CALDB_v4.7.8.html}}.
The effective exposures after the screening process are listed in table~\ref{tab:tycho_observations}.

Before performing imaging and spectroscopy, we aligned coordinates of each observation to that of the deepest observation (ObsID 10095) based on the positions of detected point sources, using the {\tt wcs\_match} and {\tt wcs\_update} tasks in CIAO.
For better statistics, we then combined two 2007 data sets in our imaging and spectral analysis.
The nine data sets in 2009 were also combined in the same way.
The total effective exposure times in 2000, 2003, 2007, 2009, and 2015 are 49~ks, 146~ks, 142~ks, 734~ks, and 147~ks, respectively. 

\begin{deluxetable}{rccc}
\tablecaption{Tycho's SNR observation log. \label{tab:tycho_observations}}
\tablecolumns{3}
\tablenum{1}
\tablewidth{0pt}
\tablehead{
\colhead{ObsID} &
\colhead{Start Date} &
\colhead{Effective Exposure} \\
\colhead{} & \colhead{(YYYY-mm-dd)} &
\colhead{(ks)} 
}
\startdata
    115  & 2000 Sep 20 & 48.91 \\
    3837 & 2003 Apr 29 & 145.6 \\
    7639 & 2007 Apr 23 & 108.87 \\ 
    8551 & 2007 Apr 26 & 33.27 \\
    10093 & 2009 Apr 13 & 118.35 \\
    10094 & 2009 Apr 18 & 89.97 \\
    10095 & 2009 Apr 23 & 173.37 \\
    10096 & 2009 Apr 27 & 105.72 \\
    10097 & 2009 Apr 11 & 107.43 \\
    10902 & 2009 Apr 15 & 39.53 \\
    10903 & 2009 Apr 17 & 23.92 \\
    10904 & 2009 Apr 13 & 34.7 \\
    10906 & 2009 May 03 & 41.12 \\
    15998 & 2015 Apr 22 & 146.98 \\
\enddata
\end{deluxetable}

\begin{figure*}
\begin{center}
 \includegraphics[width=14cm]{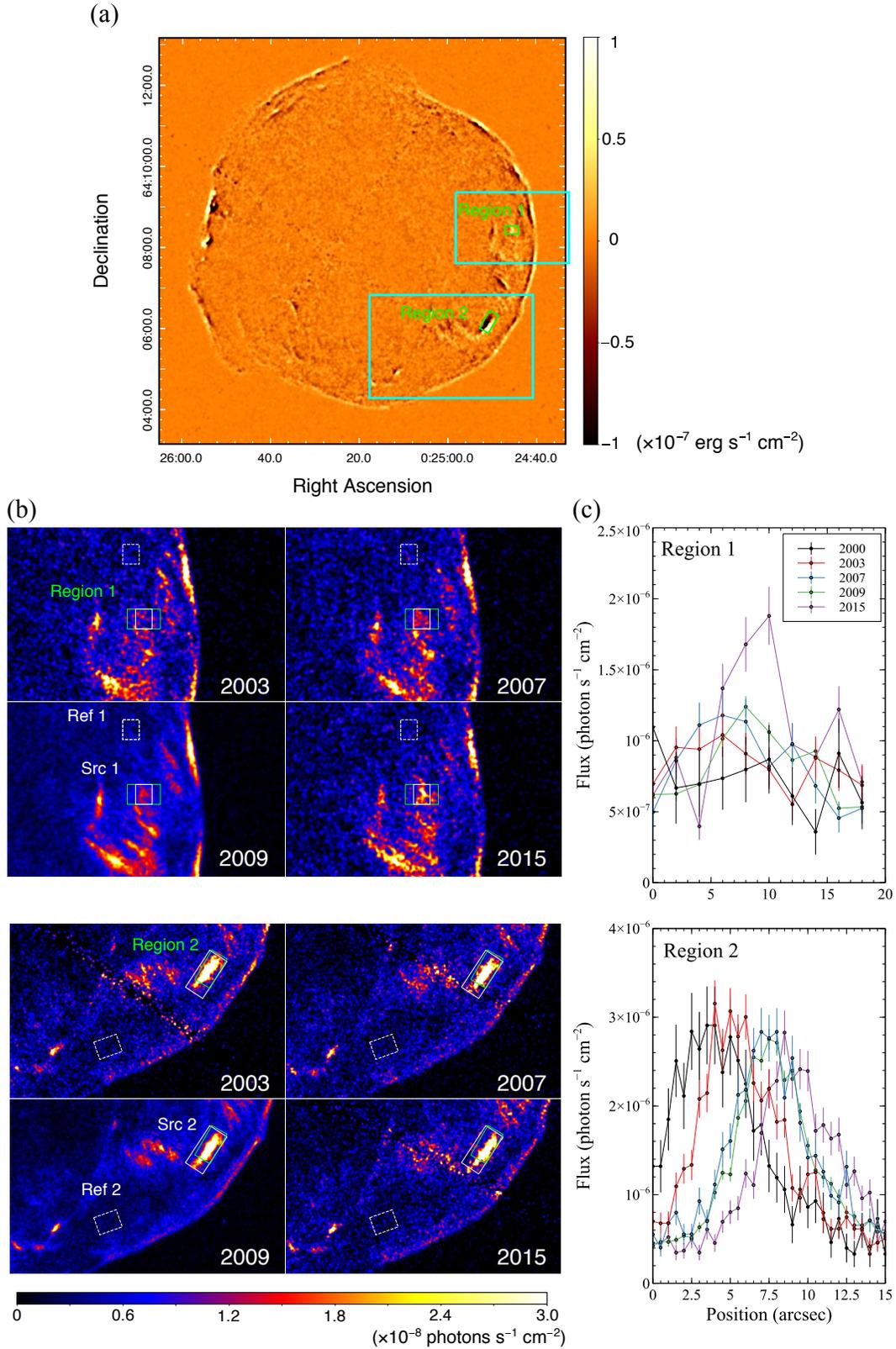} 
\end{center}
\caption{(a) Difference image between 2003 and 2015 observations in the 4.1--6.1 keV band.
The cyan boxes indicate the fields of view of Figure~\ref{fig:image}b.
Regions 1 and 2 surrounded with green solid lines are characteristic regions (see the text).
(b) Exposure-corrected X-ray (4.1--6.1 keV) images around Regions 1 ({\it top}) and 2 ({\it bottom}) in 2003, 2007, 2009, and 2015.
The regions enclosed by the green solid lines (Regions 1 and 2) are used for projections shown in Figure~\ref{fig:image}c.
The source-extraction regions (Src 1 and Src 2) and their reference regions (Ref 1 and Ref 2) are indicated by the white solid and dashed lines, respectively.
(c) Projections along azimuthal directions from Regions 1 ({\it top}) and 2 ({\it bottom}). The black, red, blue, green, and purple points correspond to the 2000, 2003, 2007, 2009, and 2015 data, respectively. 
}
\label{fig:image}
\end{figure*}

\section{Analysis and results}\label{sec:analysis}
\subsection{Imaging analysis}\label{subsec:imaging}
Figure~\ref{fig:image}a shows a difference image in the 4.1--6.1 keV band, which is dominated by nonthermal emission, made by subtracting the exposure-corrected image taken in 2003 from the image in 2015.
The expansion of the blast wave and radial proper motion of the western bright stripes are clearly visible as adjacent black-white feature pairs.
We focus on two regions around the western stripes, Regions 1 and 2, as designated in Figure~\ref{fig:image}a, where 
we found hints of X-ray flux changes from 2003 to 2015.
Region 1 shows only the white structure without any corresponding black features. 
The inner black structure in Region 2 appears larger than the outer white structure.

We present close-up views of Regions 1 and 2 in Figure~\ref{fig:image}b 
as well as projections along azimuthal directions in Figure~\ref{fig:image}c. 
The small structure in Region 1 becomes brighter in 2015.
The bright filamentary structure in Region 2 gradually becomes fainter and narrower as it moves outward.

\subsection{Spectral analysis}\label{subsec:spectra}
We perform spectral analysis of the two regions over the wide energy band of 0.5--10 keV in order to quantitatively evaluate the variability and to search for possible spectral changes.
We defined the source-extraction regions as Src 1 and Src 2 as labeled in Figure~\ref{fig:image}b.
X-ray spectra from Tycho's SNR consists of nonthermal and thermal components \citep[e.g.,][]{Hwang2002}. 
Analyzing the spectra, we found that it is difficult to constrain the thermal component because of the strong nonthermal emission in the regions Src 1 and Src 2.  
In order to accurately evaluate the contributions from the thermal components, therefore, we first analyzed the spectra extracted from reference regions, Ref 1 and Ref 2 (Figure~\ref{fig:image}b), where 
the nonthermal emission is weaker.
Backgrounds were extracted from an off-source region within the ACIS-I array. 
The spectra were binned so that each bin has at least 15 counts.

We modeled nonthermal and thermal emission, following the works by \cite{Sato2017} and \cite{Yamaguchi2017}.
We approximate the nonthermal coponent with the power-law model.
To the thermal component, we apply a two-component nonequilibrium ionization (NEI) model (vnei) in XSPEC \citep{Arnaud1996} with AtomDB version 3.0.9\footnote{\url{http://www.atomdb.org}} \citep[e.g.,][]{Foster2012}.
In addition, we added a Gaussian at $\sim 1.23~\mathrm{keV}$, because e.g., \cite{Brickhouse2000} and \cite{Audard2001} claimed that some Fe L lines from high quantum numbers are missing in AtomDB.
We employed the Tuebingen-Boulder absorption model (TBabs; \citealt{Wilms2000}) for the interstellar absorption.

The two-component NEI model represents emission from intermediate-mass elements (IMEs) and iron in the supernova ejecta.
The abundances of Mg, Si, S, Ar, Ca, and Fe are free parameters, and the Ni abundance is linked to Fe.
Since Tycho's SNR is of Type Ia origin, we fixed H, He, and N abundances to zero.
We fixed the abundances of O and Ne to the solar values with respect to C, because C has the lowest atomic number in the elements that we have in the ejecta.
We assume the common emission measure ($\equiv \frac{1}{[{\rm C}/{\rm H}]_{\odot}}\frac{1}{4 \pi d^2}\int n_{e} n_{\rm C} dV$) between the two NEI components, where $d$ is the distance to Tycho's SNR, $V$ is the volume of the emitting plasma, and $n_e$ and $n_{\rm C}$ are the number densities of electrons and C, respectively.
The ionization timescales $n_{e}t$ are fixed to the values obtained from the corresponding reference regions.

\begin{figure*}
\begin{center}
 \includegraphics[width=15.0cm]{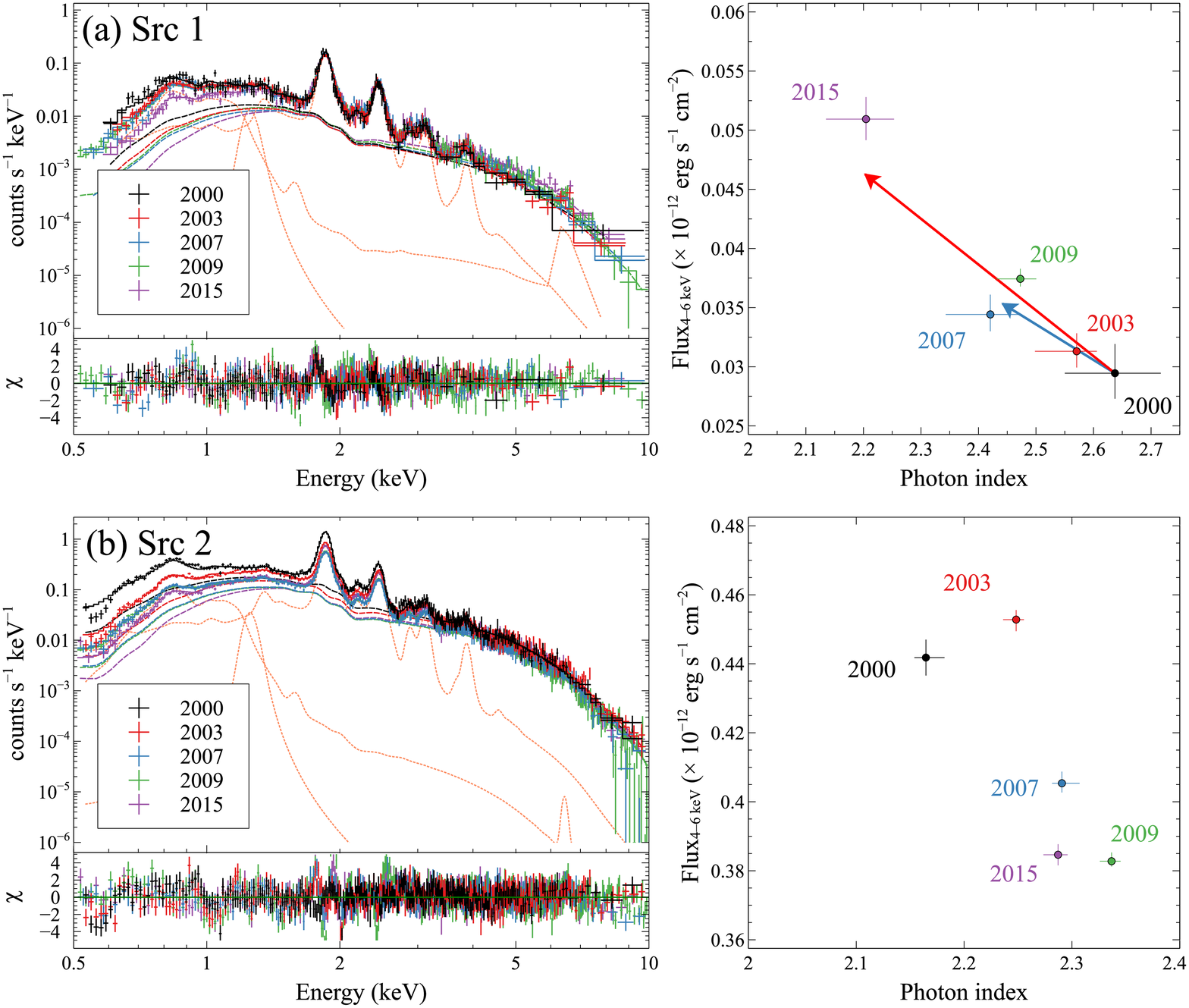} 
\end{center}
\caption{(a) {\it Left}: Spectra extracted from Src 1. The red, blue, green, and purple points indicate spectra taken in 2000, 2003, 2007, 2009, and 2015, respectively. Their best-fit models are shown as the solid lines. The dashed lines represent nonthermal components in each observation. The dotted orange lines represents the thermal components in 2003. 
{\it Right}: Relation between the photon index $\Gamma$ and the X-ray (4--6 keV) flux $F_{4\textrm{--}6~\mathrm{keV}}$. 
(b) Same as Figure~\ref{fig:spec}a but for Src 2.
The arrows represent predicted variabilities for changes in the electron cutoff energy ({\it red}) and magnetic turbulence ({\it blue}).
}
\label{fig:spec}
\end{figure*}

We performed simultaneous fitting of the spectra taken in 2000, 2003, 2007, 2009, and 2015.
Throughout the four epochs, we linked all the thermal parameters except for the normalizations.
Figure~\ref{fig:spec} shows the spectra and the best-fit models from Src 1 and Src 2.
We also list the best-fit parameters in table~\ref{tab:parameters}.
The spectra from Src 1 and Src 2 are well reproduced by the model with $\chi^2~({\rm d.o.f.})$ of 1427 (1000) and 2885 (1846), respectively.
In Figure~\ref{fig:spec}, we plot the 4--6 keV flux ($F_{4\textrm{--}6~\mathrm{keV}}$) as a function of photon index ($\Gamma$).
From 2000 to 2015, Src 1 shows brightening of $\sim 70\%$ in the 4--6 keV band and hardening of $\Delta \Gamma \sim 0.45$, 
whereas Src 2 shows darkening of $\sim 20\%$ and slight softening of $\Delta \Gamma \sim 0.2$.
The results are unchanged even when electron temperatures $kT_{e}$ are not linked among the four epochs.

We compared the observed photon index changes with expected levels of systematic errors due to effective area uncertainty of Chandra. 
According to the study by \cite{Lee2011}, the effective area uncertainty hinders determining photon indices better than 0.04, which we here take as systematic 
errors in photon indices. 
The photon index changes of Src~1 (Figure~\ref{fig:spec}a) is much larger than this number, and thus we can safely regard the result as significant. 
Most of the data points of Src~2 (Figure~\ref{fig:spec}b), on the other hand, scatter within the systematic error $\Delta \Gamma = 0.04$. 
We therefore conservatively conclude that the photon index changes that we observed in Src~2 are not significant well beyond the systematic effect 
and thus we do not further discuss them below. 
Even for Src~1, one should particularly be careful about the gradual decrease of the quantum efficiency of the ACIS in the soft band below $\sim 2~{\rm keV}$ due to 
contamination on the ACIS Optical Blocking Filters\footnote{\url{http://cxc.harvard.edu/ciao/why/acisqecontamN0010.html}}. 
We repeated the same spectral analysis as above but with an energy range restricted to $> 2~{\rm keV}$, where the contamination effect is almost negligible. 
We still obtained the same level of the time variability even from this analysis.


\section{Discussion}\label{sec:discussion}
Our imaging and spectroscopy with Chandra have revealed significant year-scale time variability of nonthermal X-ray stripes in Tycho's SNR for the first time.
One of the possible interpretations is that the flux changes are due to the changes of the magnetic field strength.
Similar variability is not clearly seen in the radio continuum band \citep[e.g.,][]{Reynoso1997,Vigh2011,Williams2016}.
Synchrotron flux is more sensitive to magnetic field strength changes in radio than in X-rays since X-ray emission is ``saturated'' due to significant 
synchrotron cooling of X-ray-emitting electrons. 
If this scenario is the case, we would see larger time variability in radio. 
We, therefore, conclude that this scenario is unlikely. 

Instead, it is more likely that the time variability is attributed to fast acceleration and synchrotron cooling loss of X-ray-emitting electrons in amplified magnetic fields. 
If acceleration and synchrotron cooling proceed in a short timescale of $\sim {\rm yr}$, the cutoff energy of electrons substantially changes in the same timescale. 
X-rays, which are radiated by electrons in the cutoff region, become brighter and harder (fainter and softer) as the electron cutoff energy increases (decreases). 
Time variability of synchrotron X-rays discovered in RX~J1713.7$-$3946 and in Cassiopeia A are actually interpreted in the same way by \cite{Uchiyama2007}
and \cite{Uchiyama2008}, respectively. 

We can estimate the magnetic field strength by comparing the observed variability timescale with timescales of particle acceleration and synchrotron cooling. 
As discussed by \cite{Uchiyama2007}, the acceleration timescale is given as 
\begin{eqnarray}
t_{\rm acc} = 4 \eta \left( \frac{\varepsilon}{\rm keV} \right)^{0.5} \left( \frac{B}{200~\mathrm{\mu G}} \right)^{-1.5} \left( \frac{v_{\rm sh}}{5000~\mathrm{km~s^{-1}}} \right)^{-2}~\mathrm{yr}, 
\label{eq:particle_acceleration}
\end{eqnarray}
where $B$ is the magnetic field strength, $\varepsilon$ is the synchrotron photon energy, $v_{\rm sh}$ is the shock velocity, and $\eta\ (\geq 1)$ is so-called ``gyrofactor.''
If the brightening and hardening observed in Region 1 are solely attributed to an increase of radiating electrons by fast acceleration, $t_{\rm acc}$ should be comparable to the observed 
variability timescale of $\sim 4~\mathrm{yr}$. Thus, the magnetic field is required to be $\sim 200~\mu{\mathrm G}$ in Region 1. 
If we interpret that X-rays in Region 2 become fainter and softer due to a decrease of radiating electrons through synchrotron cooling, the timescale of 
the variability ($\sim 4~\mathrm{yr}$) should be compared with the synchrotron cooling timescale, 
\begin{eqnarray}
t_{\rm synch} = 4 \left( \frac{B}{500~\mathrm{\mu G}} \right)^{-1.5} \left( \frac{\varepsilon}{\rm keV} \right)^{-0.5}~\mathrm{yr}. 
\label{eq:synchrotron_cooling}
\end{eqnarray}
The magnetic field needs to be as strong as $\sim 500~\mu{\mathrm G}$ in Region 2.

The strongly amplified magnetic field can affect the estimate of the maximum acceleration energy of protons in Tycho's SNR. 
\cite{Eriksen2011} estimated that the energy of accelerated protons reaches $E_\mathrm{max} = 2 \times 10^{15}~\mathrm{eV}$ assuming that the gyroradius of proton is equivalent with the half interval of the stripes.
In this estimate, the authors assumed $B = 30~\mathrm{\mu G}$, which is an upstream magnetic field strength predicted by \cite{Cassam-Chenai2007}.
Since $E_\mathrm{max}$ is proportional to $B$, our result implies that $E_\mathrm{max}$ is even higher than the estimate by \cite{Eriksen2011}, well beyond the ``knee'' of the cosmic-ray spectrum. 
On the other hand, the gamma-ray emission of Tycho's SNR, which is integrated from the whole remnant and is presumably from $\pi^0$ decay, 
has a steep spectrum particularly in the TeV range \citep{Archambault2017}. 
Particle acceleration may be proceeding up to the ``knee'' in localized regions that cannot be resolved with gamma-ray instruments. 
A smoking gun would be the detection of synchrotron radiation from secondary $e^{+}$/$e^{-}$, which are decay products of $\pi^{+}$/$\pi^{-}$ generated at the same time as $\pi^0$ by interactions between 
accelerated protons and ambient gas. 
Future hard X-ray observations with an angular resolution good enough to resolve the stripe structures are needed to give a decisive conclusion about the maximum acceleration energy. 

Compared to time-variable synchrotron X-rays in other SNRs \citep{Uchiyama2007,Uchiyama2008,Borkowski2018}, one of the new findings of ours in Tycho's SNR is that 
we observed significant spectral hardening accompanying the flux increase in Region 1. 
We tested if changes in the electron cutoff energy can explain both the flux change and spectral slope change at the same time. 
If we assume an electron spectrum in the form of Equation (24) of \cite{Zira2007}, which has a cutoff shape of $\exp[-(E/E_0)^2]$, where $E$ is the electron energy and $E_0$ is the electron cutoff energy, 
the electron cutoff energy is required to increase by 80\% to explain the spectral slope change from $\Gamma = 2.64$ to $\Gamma = 2.20$ as observed in Region 1. 
In this case, the synchrotron flux is predicted to increase by $60\%$, which is roughly consistent with the observed flux increase by $70\%$ as indicated in Figure~\ref{fig:spec}. 

We so far considered only magnetic field strengths, and implicitly assumed that the degree of the magnetic turbulence is constant as a function of time. 
However, the magnetic turbulence may grow or decay with time. 
In the following, we investigate if changes in turbulence can account for the synchrotron X-ray variability in terms of both the flux and the spectral slope. 
\cite{Zira2008} performed MHD simulations of the magnetic field amplifications due to the nonresonant streaming instability \citep{Bell2004}. 
They found that the probability density function (PDF) of the magnetic field strength can be approximated with an analytical form\footnote{There is a typo in the equation by \cite{Zira2008}. One should refer to \cite{Zira2010} for the correct equation.} of 
\begin{eqnarray}
P_B (B) = \frac{6B}{{B_\mathrm{rms}}^2} \, \exp\left( -\frac{\sqrt{6}B}{B_\mathrm{rms}} \right), 
\label{eq:pdf}
\end{eqnarray}
where $B_\mathrm{rms} = \langle B^2 \rangle^{1/2} $. 
We also tried a Gaussian \citep{Bykov2008} as well as a power-law-like function \citep{Kelner2013} to find essentially the same results. 
Let us assume an extreme case in which the PDF changes from $P_B (B) = \delta(B-B_0)$, which corresponds to a uniform magnetic field, 
to $P_B (B)$ given by Equation~(\ref{eq:pdf}), and vice versa. 
Once $P_B (B)$ is assumed, we can calculate synchrotron spectra as 
\begin{eqnarray}
\varepsilon \frac{dn}{d\varepsilon} \propto \int dB\, B\, P(B) \int p^2 \, dp \, F(p) \, R(\omega/\omega_c), 
\label{eq:sync_spec}
\end{eqnarray}
where $F(p)$ denotes the phase-space distribution of electrons, $R(x)$ is a synchrotron spectrum from a single electron in a magnetic field with chaotic directions, and $\omega_c = 1.5eBp^2/{m_e}^3 c^3$.  
We assumed here Equation (24) of \cite{Zira2007} as $F(p)$. 

\begin{figure}
\begin{center}
 \includegraphics[width=7.5cm]{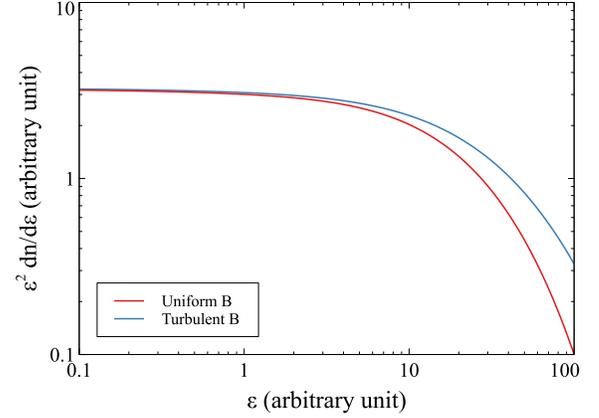} 
\end{center}
\caption{Synchrotron spectra in a uniform magnetic field ({\it red}) and a turbulent magnetic field ({\it blue}).
}
\label{fig:turbulence}
\end{figure}

In Figure~\ref{fig:turbulence}, we present synchrotron spectra in a uniform magnetic field and a turbulent magnetic field with a common electron spectrum and $B_\mathrm{rms} = B_0$ assumed. 
Synchrotron emission in the cutoff region becomes brighter and harder as the magnetic field becomes turbulent. 
Another thing to note is the flux in the lower-energy region below the cutoff is almost unchanged independent of the assumption about $P_B (B)$. 
Thus, changes in magnetic turbulence cause time variability in X-rays without any variability in radio. 
From the two spectra shown in Figure~\ref{fig:turbulence}, we found that the synchrotron flux can increase by $\sim 20\%$ as the spectrum becomes harder from $\Gamma = 2.64$ to $\Gamma = 2.44$ if the magnetic field evolves from the nonturbulent to turbulent states (Figure~\ref{fig:spec}).  
Although these numbers are somewhat smaller than those observed in Region 1, magnetic turbulence growth 
may be responsible for some portion of the variability detected with Chandra. 

As discussed above, our observational results imply strong amplification of the magnetic field and/or presence of magnetic turbulence in Regions 1 and 2 with some mechanisms. 
One of the possible and widely discussed mechanisms is instability driven by the electric current of accelerated particles \citep[e.g.,][]{Bell2001,Bell2004,Amato2006}. 
Another possible explanation is that turbulence generation and magnetic field amplification are caused by turbulent dynamo actions when the blast wave interacts with 
an ambient clumpy gas cloud. Such phenomena are studied with magnetohydrodynamical simulations by, e.g., \cite{Inoue2012} and \cite{Celli2018} and with observations 
by, e.g., \cite{Sano2015} and \cite{Okuno2018} in the case of RX~J1713.7$-$3946. 
We, however, note that, in the western region, no signature of a shock--cloud interaction is found with the expansion velocity measurements in radio \citep{Reynoso1997} 
or in X-rays \citep{Katsuda2010,Williams2016}. 
Further observational studies are needed to clarify if a shock--cloud interaction affects the magnetic field in Regions 1 and 2. 
In any scenario, one needs to explain, at the same time, both the synchrotron variability and the coherent stripes, which challenges the current understandings of 
particle acceleration.

\acknowledgments
 
We are deeply grateful to Hidetoshi Sano, Shiu-Hang Lee, Shu-ichiro Inutsuka, and Miroslav Filipovic for helpful discussion.
This work is supported by JSPS/MEXT Scientific Research Grant Nos JP25109004 (T.T. and T.G.T.), JP19H01936 (T.T.), JP26800102 (H.U.), JP19K03915 (H.U.), and JP15H02090 (T.G.T.), 
and also by the ISHIZUE grant of Kyoto University (T.T.).

\begin{longrotatetable}
\begin{deluxetable*}{llccccccccccc}
\tabletypesize{\footnotesize}
\tablecaption{Fit results of Src 1 and Src 2. \label{tab:parameters}}
\tablecolumns{13}
\tablenum{2}
\tablehead{
\colhead{Component} &
\colhead{Parameters} &
\multicolumn{5}{c}{Src 1} &
\colhead{} &
\multicolumn{5}{c}{Src 2} \\
\cline{3-7}
\cline{9-13}
\colhead{} &
\colhead{} & 
\colhead{2000} &
\colhead{2003} &
\colhead{2007} &
\colhead{2009} &
\colhead{2015} &
\colhead{} &
\colhead{2000} &
\colhead{2003} &
\colhead{2007} &
\colhead{2009} &
\colhead{2015} 
}
\startdata
    Absorption & $N_{\rm H}\tablenotemark{a} $ & \multicolumn{5}{c}{$0.74^{+0.01}_{-0.02}$} & & \multicolumn{5}{c}{$0.676^{+0.006}_{-0.002}$} \\ 
    Powerlaw & $\Gamma$ & $2.64^{+0.08}_{-0.09}$ & $2.57^{+0.04}_{-0.07}$ & $2.42^{+0.05}_{-0.08}$ & $2.47^{+0.03}_{-0.04}$ & $2.20^{+0.05}_{-0.07}$ & & $2.16^{+0.02}_{-0.01}$ & $2.25 \pm 0.01$ & $2.29^{+0.02}_{-0.01}$ & $2.34 \pm 0.01$ & $2.29 \pm 0.01$ \\
    & $F_{4\textrm{--}6~\mathrm{keV}}$\tablenotemark{b} & $0.29 \pm 0.02$ & $0.31^{+0.02}_{-0.01}$ & $0.34^{+0.02}_{-0.01}$ & $0.37 \pm 0.01$ & $0.51 \pm 0.02$ & & $4.42 \pm 0.05$ & $4.53 \pm 0.03$ & $4.05 \pm 0.03$ & $3.83^{+0.03}_{-0.02}$ & $3.85 \pm 0.03$ \\ \hline
    Ejecta & Norm.\tablenotemark{c} & $2.76^{+0.26}_{-0.13}$ & $3.21 \pm 0.18$ & $3.29^{+0.20}_{-0.30}$ & $3.33^{+0.21}_{-0.31}$ & $3.28^{+0.22}_{-0.14}$ & & $6.50^{+0.48}_{-0.42}$ & $6.60^{+0.64}_{-0.61}$ & $5.19^{+0.79}_{-0.38}$ & $5.46^{+0.38}_{-0.78}$ & $6.68^{+0.49}_{-0.89}$ \\ 
    (IME comp.) & $kT_{e}$\tablenotemark{d} & \multicolumn{5}{c}{$1.24^{+0.04}_{-0.02}$} & & \multicolumn{5}{c}{$1.823^{+0.003}_{-0.007}$} \\
    & $n_{e}t$\tablenotemark{e} & \multicolumn{5}{c}{4.74 (fixed)} & & \multicolumn{5}{c}{3.73 (fixed)} \\
    & $[{\rm Mg}/{\rm C}]/[{\rm Mg}/{\rm C}]_{\odot}$ & \multicolumn{5}{c}{$5.0^{+0.5}_{-0.3}$} & &\multicolumn{5}{c}{$6.9 \pm 0.6$} \\
    & $[{\rm Si}/{\rm C}]/[{\rm Si}/{\rm C}]_{\odot}$ & \multicolumn{5}{c}{$107 \pm 7$} & &\multicolumn{5}{c}{$190^{+27}_{-10}$}  \\
    & $[{\rm S}/{\rm C}]/[{\rm S}/{\rm C}]_{\odot}$ & \multicolumn{5}{c}{$108^{+10}_{-9}$} & &\multicolumn{5}{c}{$152^{+17}_{-8}$} \\
    & $[{\rm Ar}/{\rm C}]/[{\rm Ar}/{\rm C}]_{\odot}$ & \multicolumn{5}{c}{$134^{+10}_{-14}$} & &\multicolumn{5}{c}{$164^{+21}_{-10}$}  \\
    & $[{\rm Ca}/{\rm C}]/[{\rm Ca}/{\rm C}]_{\odot}$ & \multicolumn{5}{c}{$301^{+18}_{-45}$} & &\multicolumn{5}{c}{$307^{+73}_{-16}$} \\
    (Fe comp.) & $kT_{e}$\tablenotemark{d} & \multicolumn{5}{c}{$5.08^{+0.44}_{-0.57}$} & & \multicolumn{5}{c}{$1.20 \pm 0.01$}  \\
    & $n_{e}t$\tablenotemark{e} & \multicolumn{5}{c}{0.75 (fixed)} & & \multicolumn{5}{c}{1.29 (fixed)}  \\
    & $[{\rm Fe}/{\rm C}]/[{\rm Fe}/{\rm C}]_{\odot}$  & \multicolumn{5}{c}{$4.8 \pm 0.3$} & & \multicolumn{5}{c}{$7.2^{+0.7}_{-0.6}$} \\
    Gaussian & Norm.\tablenotemark{f} & $0.40^{+0.14}_{-0.15}$ & $0.73 \pm 0.11$ & $0.56^{+0.11}_{-0.10}$ & $0.62 \pm 0.05$ & $0.66 \pm 0.12$ & & $1.22^{+0.28}_{-0.30}$ & $1.53^{+0.20}_{-0.22}$ & $1.05^{+0.22}_{-0.23}$ & $1.10^{+0.10}_{-0.11}$ & $2.23^{+0.22}_{-0.25}$ \\ 
    & Centroid\tablenotemark{a} & \multicolumn{5}{c}{$1.26 \pm 0.01$} & & \multicolumn{5}{c}{$1.250^{+0.003}_{-0.008}$}  \\ \hline
    & $\chi^2~(\mathrm{d.o.f.})$ & \multicolumn{5}{c}{1427 (1000)} & & \multicolumn{5}{c}{2885 (1846)} \\ \hline
\enddata
\tablenotetext{a}{In units of $10^{22}~{\rm cm}^{-2}$.}
\tablenotetext{b}{In units of $10^{-13}~{\rm erg}~{\rm s}^{-1}~{\rm cm}^{-2}$}
\tablenotetext{c}{Emission measures of Fe and IME components are defined by $\int n_{e} n_{\rm C} dV / ( 4 \pi d^2 \cdot [{\rm C}/{\rm H}]_{\odot})$ in units of $10^{19}~{\rm cm}^{-5}$ and linked to each other.}
\tablenotetext{d}{In units of keV.}
\tablenotetext{e}{In units of $10^{10}~{\rm s}~{\rm cm}^{-3}$.}
\tablenotetext{f}{In units of $10^{-5}~{\rm photons}~{\rm s}^{-1}~{\rm cm}^{-2}$.}
\end{deluxetable*}
\end{longrotatetable}

\begin{thebibliography}{}


\bibitem[Amato, \& Blasi(2006)]{Amato2006} Amato, E., \& Blasi, P.\ 2006, \mnras, 371, 1251

\bibitem[Archambault et al.(2017)]{Archambault2017} Archambault, S., Archer, A., Benbow, W., et al.\ 2017, \apj, 836, 23

\bibitem[Arnaud(1996)]{Arnaud1996} Arnaud, K.~A.\ 1996, Astronomical Data Analysis Software and Systems V, 101, 17 

\bibitem[Audard et al.(2001)]{Audard2001} Audard, M., Behar, E., G{\"u}del, M., et al.\ 2001, \aap, 365, L329 

\bibitem[Baade(1945)]{Baade1945} Baade, W.\ 1945, \apj, 102, 309

\bibitem[Bamba et al.(2005)]{Bamba2005} Bamba, A., Yamazaki, R., Yoshida, T., et al.\ 2005, \apj, 621, 793

\bibitem[Bell(2004)]{Bell2004} Bell, A.~R.\ 2004, \mnras, 353, 550

\bibitem[Bell \& Lucek(2001)]{Bell2001} Bell, A.~R., \& Lucek, S.~G.\ 2001, \mnras, 321, 433

\bibitem[Borkowski et al.(2018)]{Borkowski2018} Borkowski, K.~J., Reynolds, S.~P., Williams, B.~J., et al.\ 2018, \apjl, 868, L21

\bibitem[Brickhouse et al.(2000)]{Brickhouse2000} Brickhouse, N.~S., Dupree, A.~K., Edgar, R.~J., et al.\ 2000, \apj, 530, 387

\bibitem[Bykov et al.(2008)]{Bykov2008} Bykov, A.~M., Uvarov, Y.~A., \& Ellison, D.~C.\ 2008, \apj, 689, L133

\bibitem[Bykov et al.(2011)]{Bykov2011} Bykov, A.~M., Ellison, D.~C., Osipov, S.~M., Pavlov, G.~G., \& Uvarov, Y.~A.\ 2011, \apjl, 735, L40 

\bibitem[Cassam-Chena{\"i} et al.(2007)]{Cassam-Chenai2007} Cassam-Chena{\"i}, G., Hughes, J.~P., Ballet, J., \& Decourchelle, A.\ 2007, \apj, 665, 315 

\bibitem[Celli et al.(2018)]{Celli2018} Celli, S., Morlino, G., Gabici, S., et al.\ 2018, arXiv e-prints , arXiv:1804.10579

\bibitem[Chen et al.(2017)]{Chen2017} Chen, X., Xiong, F., \& Yang, J.\ 2017, \aap, 604, A13


\bibitem[Eriksen et al.(2011)]{Eriksen2011} Eriksen, K.~A., Hughes, J.~P., Badenes, C., et al.\ 2011, \apjl, 728, L28

\bibitem[Foster et al.(2012)]{Foster2012} Foster, A.~R., Ji, L., Smith, R.~K., \& Brickhouse, N.~S.\ 2012, \apj, 756, 128

\bibitem[Garmire et al.(2003)]{Garmire2003} Garmire, G. P., Bautz, M. W., Ford, P. G., Nousek, J. A., \& Ricker, G. R., Jr.\ 2003, \procspie, 4851, 28

\bibitem[Hayato et al.(2010)]{Hayato2010} Hayato, A., Yamaguchi, H., Tamagawa, T., et al.\ 2010, \apj, 725, 894

\bibitem[Hwang et al.(2002)]{Hwang2002} Hwang, U., Decourchelle, A., Holt, S.~S., \& Petre, R.\ 2002, \apj, 581, 1101

\bibitem[Inoue et al.(2012)]{Inoue2012} Inoue, T., Yamazaki, R., Inutsuka, S.-i., \& Fukui, Y.\ 2012, \apj, 744, 71

\bibitem[Katsuda et al.(2010)]{Katsuda2010} Katsuda, S., Petre, R., Hughes, J.~P., et al.\ 2010, \apj, 709, 1387

\bibitem[Kelner et al.(2013)]{Kelner2013} Kelner, S.~R., Aharonian, F.~A., \& Khangulyan, D.\ 2013, \apj, 774, 61

\bibitem[Krause et al.(2008)]{Krause2008} Krause, O., Tanaka, M., Usuda, T., et al.\ 2008, \nat, 456, 617

\bibitem[Lee et al.(2004)]{Lee2004} Lee, J.-J., Koo, B.-C., \& Tatematsu, K.\ 2004, \apjl, 605, L113

\bibitem[Lee et al.(2011)]{Lee2011} Lee, H., Kashyap, V.~L., van Dyk, D.~A., et al.\ 2011, \apj, 731, 126

\bibitem[Okuno et al.(2018)]{Okuno2018} Okuno, T., Tanaka, T., Uchida, H., Matsumura, H., \& Tsuru, G.~T.\ 2018, \pasj, 70, 77





\bibitem[Reynoso et al.(1997)]{Reynoso1997} Reynoso, E.~M., Moffett, D.~A., Goss, W.~M., et al.\ 1997, \apj, 491, 816

\bibitem[Ruiz-Lapuente(2004)]{Ruiz-Lapuente2004} Ruiz-Lapuente, P.\ 2004, \apj, 612, 357

\bibitem[Sano et al.(2015)]{Sano2015} Sano, H., Fukuda, T., Yoshiike, S., et al.\ 2015, \apj, 799, 175

\bibitem[Sato \& Hughes(2017)]{Sato2017} Sato, T., \& Hughes, J.~P.\ 2017, \apj, 840, 112


\bibitem[Tian \& Leahy(2011)]{Tian2011} Tian, W.~W., \& Leahy, D.~A.\ 2011, \apjl, 729, L15


\bibitem[Uchiyama et al.(2007)]{Uchiyama2007} Uchiyama, Y., Aharonian, F.~A., Tanaka, T., Takahashi, T., \& Maeda, Y.\ 2007, \nat, 449, 576 

\bibitem[Uchiyama \& Aharonian(2008)]{Uchiyama2008} Uchiyama, Y., \& Aharonian, F.~A.\ 2008, \apjl, 677, L105


\bibitem[Vigh et al.(2011)]{Vigh2011} Vigh, C.~D., Vel{\'a}zquez, P.~F., G{\'o}mez, D.~O., et al.\ 2011, \apj, 727, 32 

\bibitem[Williams et al.(2016)]{Williams2016} Williams, B.~J., Chomiuk, L., Hewitt, J.~W., et al.\ 2016, \apjl, 823, L32

\bibitem[Wilms et al.(2000)]{Wilms2000} Wilms, J., Allen, A., \& McCray, R.\ 2000, \apj, 542, 914

\bibitem[Yamaguchi et al.(2017)]{Yamaguchi2017} Yamaguchi, H., Hughes, J.~P., Badenes, C., et al.\ 2017, \apj, 834, 124

\bibitem[Zhou et al.(2016)]{Zhou2016} Zhou, P., Chen, Y., Zhang, Z.-Y., et al.\ 2016, \apj, 826, 34

\bibitem[Zirakashvili, \& Aharonian(2010)]{Zira2010} Zirakashvili, V.~N., \& Aharonian, F.~A.\ 2010, \apj, 708, 965

\bibitem[Zirakashvili \& Ptuskin(2008)]{Zira2008} Zirakashvili, V.~N., \& Ptuskin, V.~S.\ 2008, \apj, 678, 939

\bibitem[Zirakashvili \& Aharonian(2007)]{Zira2007} Zirakashvili, V.~N., \& Aharonian, F.\ 2007, \aap, 465, 695

\end{thebibliography}
\end{document}